\documentclass[journal]{IEEEtran}

\usepackage{graphicx}
\usepackage{float}
\usepackage{dirtytalk}
\usepackage{amsmath}
\usepackage{algorithm} 
\usepackage{algpseudocode} 
\usepackage[dvipsnames]{xcolor}
\usepackage{array}
\usepackage{multirow}
\usepackage{footmisc}
\usepackage{hyperref}

\ifCLASSINFOpdf

\else
  
\fi

\begin{document}

\title{A Hybrid Method for Condition Monitoring and Fault Diagnosis of Rolling Bearings With Low System Delay}

\author{Sulaiman~Aburakhia,~\IEEEmembership{Student Member,~IEEE,}
        Ryan~Myers,
      and~Abdallah~Shami,~\IEEEmembership{Senior~Memeber,~IEEE}%

\thanks{S. Aburakhia and A. Shami are with the Department of
Electrical and Computer Engineering, Western University, %London, ON
N6A 3K7, Canada (e-mail: saburakh@uwo.ca; abdallah.shami@uwo.ca). R. Myers is with the National Research Council Canada, London, ON N6G 4X8, Canada (e-mail: ryan.myers@nrc-cnrc.gc.ca)}}% <-this % stops a space

\markboth{}%
{Shell \MakeLowercase{\textit{et al.}}: Bare Demo of IEEEtran.cls for IEEE Journals}

\maketitle

\begin{abstract}
Vibration-based condition monitoring techniques are commonly used to detect and diagnose failures of rolling bearings. Accuracy and delay in detecting and diagnosing different types of failures are the main performance measures in condition monitoring. Achieving high accuracy with low delay improves system reliability and prevents catastrophic equipment failure. Further, delay is crucial to remote condition monitoring and time-sensitive industrial applications. While most of the proposed methods focus on accuracy, slight attention has been paid to addressing the delay introduced in the condition monitoring process. In this paper, we attempt to bridge this gap and propose a hybrid method for vibration-based condition monitoring and fault diagnosis of rolling bearings that outperforms previous methods in terms of accuracy and delay. Specifically, we address the overall delay in vibration-based condition monitoring systems and introduce the concept of system delay to assess it. Then, we present the proposed method for condition monitoring. It uses Wavelet Packet Transform (WPT) and Fourier analysis to decompose short-duration input segments of the vibration signal into elementary waveforms and obtain their spectral contents. Accordingly, energy concentration in the spectral components---caused by defect-induced transient vibrations---is utilized to extract a small number of features with high discriminative capabilities. Consequently, Bayesian optimization-based Random Forest  (RF) algorithm is used to classify healthy and faulty operating conditions under varying motor speeds. The experimental results show that the proposed method can achieve high accuracy with low system delay.   

\end{abstract}

\begin{IEEEkeywords}
Wavelet Packet Transform (WPT), fault diagnosis, condition monitoring, rolling bearing, system delay
\end{IEEEkeywords}

\IEEEpeerreviewmaketitle

\begin{figure*}[t]
\centerline{\includegraphics[width=\textwidth]{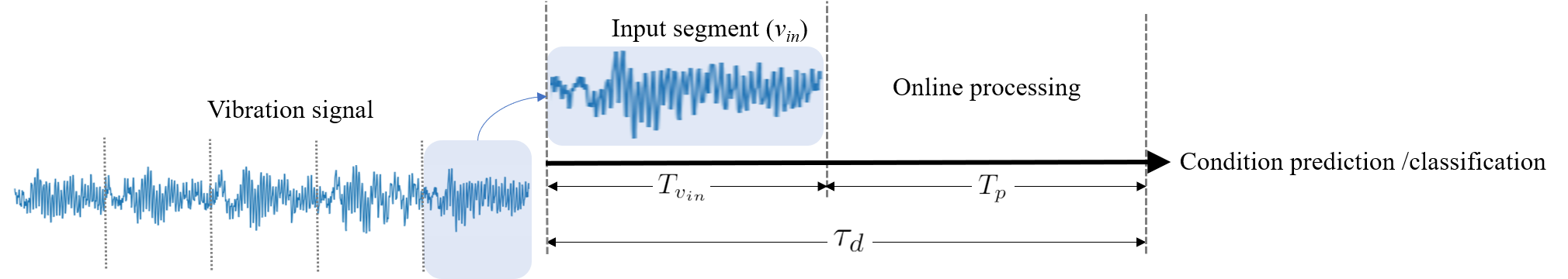}}
\caption{System delay of vibration-based condition monitoring.}
\label{fig1}
\end{figure*}

\section{Introduction}

\IEEEPARstart{T}{he} ongoing automation of industrial manufacturing---commonly referred to as Industry 4.0, smart manufacturing, or Industrial Internet of Things (IIoT)---has many potential benefits for industry and consumers. Managing maintenance is essential in smart manufacturing; effective and efficient maintenance sustains equipment availability and reliability, which in turn, ensuring safety, productivity, quality, and on-time delivery. Maintenance management strategies are generally categorized under three main categories: Corrective Maintenance (CM), Preventive Maintenance (PvM), and Predictive Maintenance (PdM). In CM, corrective actions take place after equipment failure. Despite its simplicity, CM is very costly as it involves shutting down the production process and replacing parts. PvM is a time-based scheduled strategy where maintenance is regularly performed.  It is an effective strategy as it prevents equipment failure. However, PvM involves unnecessary routine preventive actions, which increases maintenance costs. PdM is based on the continuous monitoring of the equipment's condition, and actions for maintenance are predicted based on the equipment's actual condition. This is achieved by adopting predictive approaches to monitor the equipment's functional-process integrity continuously.\\

Data-driven-based approaches have been widely adopted in recent years for PdM applications. Generally, data-driven-based PdM can be summarized under two main steps: The first step involves feature extraction, where discriminative features are extracted from equipment-related data. The second step involves utilizing the extracted features to classify or predict normal and abnormal operational conditions accurately. Hence, it is essential to extract features that can represent integrity of the operational process with high sensitivity to any changes within the process.\\

In terms of performance evaluation, accuracy and complexity are commonly used to evaluate the performance of a PdM system. Higher accuracy improves system reliability and less complexity---in terms of memory and processing time---relaxes the computational requirements of the system. Besides accuracy and complexity, delay in predicting the current condition of the running equipment is another important aspect of system performance. Delay is crucial to time-sensitive industrial applications and remote condition monitoring in IIoT as it directly influences the end-to-end latency. Further, it is a critical factor for the early detection of failures. As a real-life example, in one case \cite{slacey}, the failure of a large gearbox caused a three-week shutdown, and extensive repair costs are typically €50,000 to €100,000. After implementing a condition-monitoring system, early detection of the gearbox failure resulted in a repair cost of €5000, saving the customer at least €27,000. Moreover, the company avoided lost production, amounting to around €6000/hour. Such real-life cases demonstrate the important role of early failure detection in condition monitoring systems as it helps to detect the failure in its early stages and prevents catastrophic equipment failure. Hence, it ensures workplace safety and productivity, and reduces maintenance costs. While most of the proposed work in the literature focuses on accuracy and complexity in assessing the performance of condition monitoring systems, little attention has been paid to addressing the overall delay in the condition monitoring process.\\ 

This paper attempts to bridge this gap and introduces a base for overall delay analysis in vibration-based condition monitoring systems. Specifically, the paper defines and analyzes the overall delay in vibration-based condition monitoring and proposes a hybrid wavelet-based method for vibration-based condition monitoring and fault diagnosis of rolling bearings with low delay. The following are the main contributions of the paper:
\begin{itemize}
 \item Analyzes the overall delay in vibration-based condition monitoring and introduces the concept of system delay to assess it.
 
 \item Proposes a hybrid wavelet-based method with reduced system delay for vibration-based condition monitoring and fault diagnosis of rolling bearings under varying motor speeds. The proposed method has a high sensitivity to fault-related transients with relatively short durations of input vibration segments. 
 \item The proposed method combines WPT and FFT to decompose the input vibration segment into elementary waveforms with high time-frequency localization and obtain spectral components of these waveforms to achieve high sensitivity with short durations of the vibration signal.
  \item  Accordingly, the proposed method introduces a new technique to extract fault-sensitive features from spectral components of the elementary waveforms. Specifically, the proposed method utilizes high concentration in the spectral energy caused by defect-induced transient vibrations to select the most dominant frequency components (\textit{i.e.}, frequencies with the highest power levels).
  \item The proposed method allows controlling the size of extracted features through selection of number of decomposition levels and number of selected dominant frequencies, which helps to reduce redundancy in the extracted features. Moreover, this flexibility allows to design and adapt the proposed method according to various operational situations to meet specific application requirements in terms of accuracy and complexity.

\end{itemize}

The rest of the paper is outlined as follows: The next section provides the theoretical foundation. Section 3 presents a literature review on the application of wavelets in condition monitoring and summarizes related work. Section 4 introduces the proposed method for vibration-based condition monitoring and fault diagnosis of rolling bearings. Section 5 presents the datasets and the experimental setup for performance evaluation, while Section 6 discusses the results. The paper is finally concluded in Section 7.

\section{Theoretical Background}
\subsection{System Delay}
\textbf{\textit{The system delay $\tau_d$ of a vibration-based condition monitoring system can be defined as the time the system takes to acquire input vibration segment and classify or predict the operational condition of the current state $S_{c}$ of the equipment}}. In vibration-based monitoring, the current state $S_{c}$ is represented by the input segment $v_{in}$ of the generated vibration signal as illustrated in Fig.\ref{fig1}. Accordingly, the system delay is the sum of the time duration of input segment $T_{v_{in}}$ and the online processing time $T_p$. The system delay $\tau_d$ can be formulated mathematically as follows;
\begin{equation}
\tau_d = T_{v_{in}}+T_p
\end{equation}
The time duration $T_{v_{in}}$ of the input segment $v_{in}$ depends on number of data points in the segment. It can be expressed as:

\begin{equation}
T_{v_{in}}=\frac{N_o}{f{s}}\hspace{0.5cm} (seconds)
\end{equation}
where $N_{o}$ is number of data points in $v_{in}$ and $f_{s}$ is the sampling frequency in samples per second. Online processing time $T_p$ is algorithm-dependent; it involves two tasks, feature extraction (including pre-processing) and condition prediction/classification. Hence, online processing time $T_p$ can be generally viewed as a function of the number of data points $N_o$ in the input segment $v_{in}$, the size of extracted features $S$, and available computing resources $R_{comp}$, \textit{i.e.,}
\begin{equation}
T_p = f(N_o, S, R_{comp})
\end{equation}

Based on the above formulation, designing a condition-monitoring system with low system delay $\tau_d$ involves three main requirements:
    
\begin{enumerate}
\item Extracting features of high sensitivity to fault-related transients to improve system accuracy.
\item Extracting features of small size $S$.
\item Utilizing input vibration segments of relatively short time duration $T_{v_{in}}$ or equivalently, of small number of data points $N_o$.
\end{enumerate}

Accordingly, considering a fixed computing resources $R_{comp}$, a reliable design of vibration-based condition monitoring systems with low system delay would address the following two conditions:
\begin{equation}
\label{4}
minimize(N_o, S)
\end{equation}
and
\begin{equation}
\label{5}
maximize(Accuracy).
\end{equation}

For a given vibration-based condition monitoring system, its parameters can be tuned empirically to achieve a good trade-off between Eq. \eqref{4} and Eq. \eqref{5}. Furthermore, the optimal parameters may be obtained by formulating a multi-objective optimization problem and using Eq. \eqref{4} and Eq. \eqref{5} as objective functions.

\subsection{Wavelet Packet Transform}
\begin{figure*}[t]
\centerline{\includegraphics[width=\textwidth]{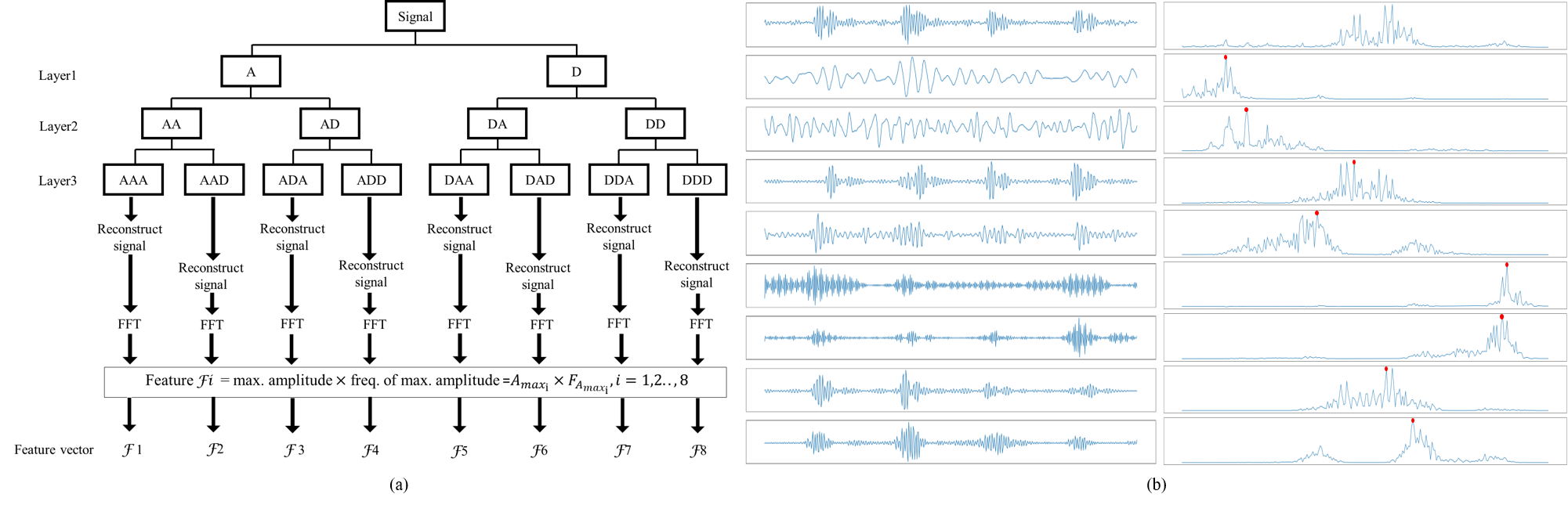}}
\caption{(a) Proposed feature extraction using 3-level WPT. (b) From top to bottom, vibration signal and its WPT-based elementary waveforms "on the left" and corresponding signal spectrum "on the right".}
\label{fig2}
\end{figure*}

Wavelet Packet Transform (WPT) offers a reliable approach to engineering  a few features from vibration signals with a high discriminative degree. Specifically, WPT is very useful in decomposing the input signal using scaled versions of a base wavelet function into elementary waveforms with high localization in time and frequency. Thus, WPT has a high time–frequency resolution, allowing to capture high-frequency transient components in the signal. 
Fig. \ref{fig2}.(a) shows the flowchart of a 3-level wavelet-packet decomposition process. The input signal is decomposed into lower and higher frequency sub-bands at each level. The
output of the decomposition process in each level is either the  approximation wavelet coefficients "$A$" associated with lower frequency bands or detail  wavelet  coefficients "$D$" associated with higher frequency bands. Accordingly, the elementary waveforms of the signal can be reconstructed using the individual wavelet coefficients. Hence, a highly discriminative feature can be extracted  from each reconstructed waveform or its corresponding wavelet coefficients. Numerically, WPT is implemented through the iterative decomposition of the signal by a series of low-pass filters $h(k)$ and high-pass filters $g(k)$ as follows \cite{xding}:
\begin{equation}
w_{i+1}^{2s}(t)=\sum_{k}h(k)w_{i}^{s}(2t-k),
\end{equation}
\begin{equation}
w_{i+1}^{2s+1}(t)=\sum_{k}g(k)w_{i}^{s+1}(2t-k),
\end{equation}
where $k=1, 2,..$ is decomposition level. The raw signal $x(t)$ is $w_{0}^{1}(k)$ and  $w_{i}^{s}(k)$ is the wavelet decomposition coefficients of node $s$ at level $i$. $w_{i+1}^{2s}(k)$ and $w_{i+1}^{2s+1}(k)$
are the wavelet coefficients of nodes $2s$ and $2s+1$ at level $i+1$, respectively, which correspond to the approximation coefficients and detail coefficients.  $h(k)$ and $g(k)$ are low pass and high pass filters which are related to the base wavelet function.

\section{Literature Review And Related Work}

Wavelet-based decomposition has been widely adopted in many applications as it offers a flexible tool for analyzing signals with a high time-frequency resolution. During the last two decades, wavelet-based approaches gained high popularity among researchers in the field of machinery diagnostics. Most of the proposed wavelet-based approaches in the literature are combined with other techniques to enhance the discriminative power of the extracted features. Accordingly, wavelet-based approaches can be generally grouped under four main categories: entropy-based \cite{cliangcl}, energy-based \cite{xding}\cite{mgom}, spectral-based \cite{xhuang22}-\cite{llia}, and statistical based\cite{wfanq232}. Spectral and statistical approaches are commonly applied to the elementary waveforms reconstructed from wavelet decomposition coefficients; entropy and energy approaches are either applied to the wavelet decomposition coefficients or to the reconstructed elementary waveforms. In \cite{mros}, it was shown that using the reconstructed waveforms leads to a better fault diagnosis than wavelet packet coefficients, especially with a low Signal-to-Noise Ratio (SNR). Entropy is a measure of uncertainty; it is used to measure the irregularity of time-series data. It is known that vibration signals of healthy bearings have high irregularity compared to faulty bearings \cite{yli}. Hence, combining wavelet decomposition and entropy measures helps quantify the degree of irregularity of a vibration signal with a high time-frequency localization. Energy-based approaches utilize the changes in energy contents of wavelet sub-bands to extract fault-related features. However, energy content has low sensitivity to incipient faults since the change in energy content is not significant at the early stages of the fault \cite{ryan}. This low sensitivity to the incipient faults can be improved by integrating entropy measures with wavelet energy-based approaches \cite{ryan2}. This approach---commonly known as wavelet energy entropy---is based on applying entropy measures to wavelet energy contents. Spectral-based approaches use spectral characteristics of the reconstructed elementary waveforms to extract the features; common techniques to extract the fault-related spectral characteristics from wavelet-decomposed  vibration signals involve Fourier transform \cite{xhuang22}, envelope spectrum \cite{lsong}\cite{hlit}, spectral kurtosis \cite{xwang4343}, and spectral entropy \cite{llia}. Statistical-based approaches rely on time-domain properties of reconstructed waveforms such as skewness, kurtosis, Root-Mean-Square (RMS), and crest factor for feature creation. Time-domain properties have low robustness to noise compared to the other approaches. Moreover, they generally have low sensitivity to defect-induced transient, which requires relativity long input segments of the vibration signal.\\ 

Recently, deep learning has been widely used in fault diagnosis and condition monitoring applications. The application of wavelet-based approaches within deep learning-based fault diagnosis falls mainly under two areas: 1) Transforming 1-D vibration signals into 2-D time-frequency input data \cite{pliangc}\cite{zshij}. 2) Utilizing the wavelet concept to design network elements such as  activation function \cite{hshaom} and convolution kernel \cite{mliaoc353}. The main idea of deep learning-wavelet-based approaches is based on utilizing wavelets' powerful time-frequency localization characteristics to extract highly disseminative features and improve the learning process.\\

As mentioned earlier, most of the proposed vibration-based condition monitoring approaches focus on the accuracy of fault detection and system complexity, while less attention has been paid to addressing the delay of the system in terms of input signal length and size of features vector. In \cite{zmezni}, Empirical Mode Decomposition (EMD) was used to decompose the vibration signals. Accordingly, an energy-based analysis is performed to select the most energized components. Further, a statistical analysis was conducted to remove redundant data. In \cite{hahmed}, a combined compressive sampling feature ranking method is proposed to learn fewer features from vibration data optimally. In \cite{yliuf233}, kurtosis value and Fisher discrimination criterion were used to screen wavelet coefficients and extract fault-transient features to enhance the discriminative capability. In \cite{cliangcl}, an entropy-wavelet-based approach was proposed to extract the features from vibration signals, and the Laplacian Score (LS) method was used to rank and select the features. In these approaches, attention is paid to reducing feature size only without considering the duration of the input signal or the processing time. In other words, in these approaches, the system is designed to reduce the dimensionality of the extracted features rather than extracting a small number of features. Although dimensionality reduction helps reduce memory requirements and improve the training process, it increases online processing time, increasing system delay. In \cite{mJalayer}, a deep learning-based method for condition monitoring is proposed, and an input length sensitivity analysis was performed. It was shown that 100\% accuracy could be achieved with shorter input lengths compared to other approaches. However, the feature engineering process utilizes vibration signals from two accelerometer sensors and involves Fast Fourier Transform (FFT), Continuous Wavelet Transform (CWT), and statistical proprieties of the raw signals to construct the features. Thus, resulting in large size features vector. This paper addresses the system delay from both aspects: the input signal's duration and the extracted features' size. Accordingly, we propose a hybrid WPT-FFT method with low system delay for vibration-based condition monitoring and fault diagnosis of rolling bearing. The next section introduces the proposed method.

\section{Proposed method for WPT-FFT Features Extraction}

As mentioned earlier, the design of a condition-monitoring system with low delay has three main requirements:
    
\begin{enumerate}
\item Extracting features of high sensitivity to fault-related transients.
\item Extracting features of small and controllable size.
\item Utilizing input vibration segments of short time durations.
\end{enumerate}
In rolling bearings, the contact between the rolling elements and the defective spot results in repetitive impulse periods. These repetitive impulses appear as a dominant frequency in the spectrum of the vibration signal \cite{wcaes}. Fourier-based analysis such as FFT provides an effective way to transform the vibration signal to the frequency domain and reveal its spectral contents. However, these contents have no time resolution, and hence, Fourier analysis provides a poor representation of signals that are highly localized in time. On the other hand, WPT  provides a very efficient time-frequency analysis of vibration signals with a good capture of fault-related transients \cite{Gyen}. Based on this concept, we propose a hybrid WPT-FFT method to fulfill the aforementioned requirements and extract a small number of highly discriminative features from short-duration vibration signals. The first step involves decomposing the input signal using $k$-level WPT. Accordingly, $2^k$ elementary waveforms of lower and higher frequency sub-bands are reconstructed from individual wavelet coefficients. In the second step, FFT is applied to the resultant waveforms to obtain the spectral contents of each waveform. Hence, a feature vector of size  $S=1\times(m\times2^k)$ can be reconstructed by utilizing the first $m$ dominant frequency components in the spectrum of each waveform. Fig.\ref{fig2} illustrates the proposed method with $k=3$ and $m=1$. As shown in Fig. \ref{fig2}.(a), after decomposing the original vibration signal and reconstructing the elementary waveforms, FFT is applied to the reconstructed waveforms to obtain and select the first dominant frequency (\textit{i.e.} the frequency with the maximum amplitude) of each waveform. Fig. \ref{fig2}.(b). shows vibration signal and elementary waveforms along with their spectral contents, where the first dominant frequency is shown as a red dot on the spectrum. After selecting the first dominant frequency, the feature of each waveform is calculated according to the below formula:
\begin{equation}
\label{1}
\mathcal{F}_i = A_{{max}_i} \times F_{A_{{max}_i}},
\end{equation}
where, $i = 1, 2,...2^k$, $A_{{max}_i}$, and $F_{A_{{max}_i}}$ are the spectral maximum amplitude of the \textit{i}th waveform and its corresponding frequency, respectively.
For $m > 1$, the first $m$ dominant  spectral amplitudes of each waveform  and their corresponding frequencies are selected, and the feature of each waveform is calculated according to the below formula:
\begin{equation}
\label{2}
\mathcal{F}_i = A1_{{max}_i} \times F_{A1_{{max}_i}},...,Am_{{max}_i} \times F_{Am_{{max}_i}}. 
\end{equation}

\begin{algorithm}
	\caption{ Proposed WPT-FFT Features Extraction} 
	\begin{algorithmic}
		\State \textit{Input:} $x:$ input segment of vibration signal of length $N_o$ data points
		\State \textit{Parameters:} $k$ = WPT decomposition level, $m$ =  number of most dominant  frequency components to be selected.
		\State \textit{Output:} $\mathcal{F}[s], s= 1,...,S:$ features vector of size $S=1\times(m\times2^k)$
		\State \textbf{Start:}
	    \State Set $k$ and $m$
		\State $s= 1$
		\State Obtain  $2^k$ elementary waveforms $w[i], i= 1,...,2^k$ of $x$ using $k$-level WPT.
		\For {$i=1,\ldots,2^k$}
		\State Compute FFT of $w[i]$ and obtain:
		\State\ $A_{FFT}:$ vector of FFT spectrum amplitudes
		\State $F_{A_{FFT}}:$ vector of FFT spectrum frequencies
			\For {$ii=1,\ldots,m$}
				\State $A_{max} = max[A_{FFT}]$: max. spectrum's amplitude
				\State $F_{A_{max}} =F_{A_{FFT}}[index(A_{max})]$: Freq. of max. spectrum's amplitude
				\State Compute feature $f= A_{max} \times F_{A_{max}}$
				\State $\mathcal{F}[s] = f$
				\State $ s= s+1$
				\State Remove $A_{max}$ from $A_{FFT}$
				\State Remove $F_{A_{max}}$ from $F_{A_{FFT}}$
			\EndFor
		\EndFor
		\State \textbf{End}
	\end{algorithmic}
	\label{alg:alg1}
\end{algorithm}

Algorithm \ref{alg:alg1} shows the pseudo-code of the proposed method. By decomposing the vibration signal into elementary waveforms and obtaining the dominant frequencies of each waveform, features with high sensitivity to fault-related transients can be calculated from these frequencies according to Eq. \eqref{1} or Eq. \eqref{2}. This has the effect of applying frequency-selective filters to the spectrum of each waveform so that only dominant frequencies are selected and then scaled by their amplitude values to increase their discriminative capabilities. This way, the remaining ineffectual spectral contents are filtered out, the redundancy is reduced, and the sensitivity to faults is improved. Moreover, since the size $S$ of the features vector equals to $1\times(m\times2^k)$, the proposed method allows controlling size of the extracted features by choosing the number of WPT decomposition levels $k$ and the number of selected dominant frequency components $m$. This, in turn, provides high flexibility and scalability in the proposed method when addressing various operational situations. Moreover, it helps reduce redundancy and meet specific application requirements in terms of accuracy and complexity through proper tuning of ($k, m$) parameters.

\subsection{Selection of Base Wavelet and Decomposition Level}
For vibration-based condition monitoring, the proper base wavelet is the wavelet that is highly correlated with defect-induced transient vibrations. This leads to high energy concentrated at the corresponding wavelet coefficients only. Due to this high energy concentration, the entropy of the energy distribution of the wavelet coefficients will be minimized. The same concept can be used to determine the appropriate decomposition level $k$ since the proper decomposition level will have the highest energy concentration and minimum entropy. Accordingly, the proper base wavelet and decomposition level can be selected by evaluating the energy and entropy values of the wavelet coefficients for different base wavelets and decomposition levels. The energy of wavelet coefficients at decomposition level $i$ can be expressed as:

\begin{equation}
E_{i}=\sum_{k=1}^{L}\left|w_{i}(k)\right|^2,
\end{equation}
where $w_{i}(k)$ are the wavelet coefficients at level $i$ and $L$ is the total number of the coefficients in that level. The  entropy  of  the  energy distribution of the wavelet coefficients can be obtained by the Shannon entropy formula:

\begin{equation}
Entropy_i=\sum_{k=1}^{L}p_ilog_2p_i,
\end{equation}

\begin{equation}
p_i=\frac{\left|w_{i}(k)\right|^2}{E_{i}}
\end{equation}
where $p_i$ is the energy probability distribution of the wavelet coefficients. The energy-to-entropy ratio proposed in \cite{ryan2} relates energy and entropy values of the wavelet coefficients and provides a useful quantitative measure to select the appropriate base wavelet and decomposition level. The energy-to-entropy ratio is expressed as \cite{ryan2}:

\begin{equation}
R(s)=\frac{E_i}{Entropy_i}
\end{equation}

Thus, the higher $R(s)$, the more appropriate base wavelet and/or decomposition level to select.

\subsection{Complexity Analysis}
Considering fixed computational resources, the system delay $\tau_d$ of vibration-based condition monitoring depends entirely on three factors:

\begin{itemize}
    \item duration of the input segment $T_{v_{in}}$,
    \item algorithm-based computations,
    \item and size of features vector $S$.
\end{itemize}
For the proposed algorithm, these factors include: number of data points $N_o$ in the input segment $v_{in}$, decomposition level $k$, WPT transform, reconstruction of $2^k$ elementary waveforms, $2^k$ FFT transforms, and number of selected dominant frequencies $m$. Accordingly, the complexity of the proposed algorithm can be analyzed as follows:

\begin{itemize}
    \item Complexity of WPT is $O(N_o\log{}N_o)$.
    \item Let $c$ represents complexity of reconstructing one elementary waveform, then complexity of reconstructing $2^k$ elementary waveforms is $O(2^kc)$.
    \item Complexity of FFT computations is $O(2^kN_o\log{}N_o)$.
    \item Complexity of feature vector computations is $O(m2^k)$.
\end{itemize}

Thus, the system delay and complexity grow as a function of ($N_o, 2^k, m$).

\section{Performance  Evaluation}
Performance of the proposed method is evaluated on the Case Western Reserve University (CWRU) bearing dataset \cite{CWRU}, the Paderborn University (PU) bearing dataset \cite{paderborn}, and the University of Ottawa (uOttawa) bearing dataset \cite{uoo}. These datasets are selected to simulate various practical situations regarding defect types, rotational speed conditions, and data sampling rates. Specifically, the CWRU dataset is very useful in benchmarking as it is widely used in the literature. However, it doesn't include combined defects and lacks real damages since faults were artificially generated in the bearings. Vibration signals of CWRU have a sampling rate of 12 KHz. In contrast to CWRU, the PU dataset has real bearing damages with combined defects. Vibration signals were sampled at 64 KHz. uOttawa dataset was released in 2019; experiments were conducted using a machinery fault simulator. The main aspect of this dataset is that it has different healthy and faulty conditions with combined defects and under time-varying rotational speed conditions. Thus, in contrast to CWRU and PU datasets, this dataset has time-varying rotational speeds within the same measurement. Further, vibration signals have a high sampling rate of 200 KHz.

\subsection{Experimental Setup}

\subsubsection{CWRU Bearing Dataset}
In the CWRU dataset, experiments are conducted using a 2 hp (horse power) Reliance Electric motor and vibration data was collected using accelerometers. Faults ranging from 0.007 to 0.040 inches in diameter were introduced separately at the Inner Raceway (IR), ball, and Outer Raceway (OR). Faulted bearings were reinstalled into the test motor, and vibration data was recorded at motor loads of 0 to 3 horsepower (motor speeds of 1,720 to 1,797 rpm). Digital data was collected at 12,000 samples per second. The dataset used in this paper consists of vibration signals generated by the accelerometers placed at the drive end of the motor housing. Table \ref{dataset_table} shows operational condition, fault diameter, and motor speed (rpm) of these vibration signals. According to operational conditions and fault diameter, faulty operational conditions are classified into nine classes. Hence, the dataset consists of ten classes, one healthy operational class and nine faulty operational classes. Vibration signals are divided into input segments of 300, 600, 1,200, and 2,400 data points to evaluate the proposed method at different time durations of input segments. With a sampling rate of 12,000 samples per second, this gives input segments with $T_{v_{in}}$ values of 0.025 seconds, 0.5 seconds, 0.1 seconds, and 0.2 seconds, respectively. For each time duration, values of $k=2,3,5$ and $m= 1,2,3$ are used to decompose the input segment and construct feature vectors. The aim here is to examine the impact of systems parameters ($k, m$) on the performance and to assess the proposed method at different sizes of features vector.

\subsubsection{PU Bearing Dataset}
In the PU dataset, experiments are conducted using 425 W Permanent Magnet Synchronous Motor (PMSM). The dataset used in this paper is based on measurements conducted at n = 1,500 rpm with a load torque of M = 0.7 Nm and a radial force on the bearing of F = 1,000 N. Vibration signals were recorded with a sampling rate of 64,000 samples per second by measuring the acceleration of the bearing housing at the adapter at the top end of the rolling bearing module. Regarding bearing defects, the PU dataset includes artificially generated and accelerated-lifetime defects. In this paper, only accelerated-lifetime defects are used. Accordingly, the dataset has four classes: one healthy class and three faulty classes according to fault type, as shown in Table \ref{dataset_table}. To evaluate the proposed method at different time durations of input segments, each vibration signal is divided into input segments of 1,600 and 12,800 data points. With a sampling rate of 64,000 samples per second, this gives input segments with $T_{v_{in}}$ values of 0.025 seconds and 0.2 seconds, respectively. For each time duration, values of $k=3,5$ and $m= 3,5,7$ are used to decompose the input segment and construct feature vectors.\

\subsubsection{University of Ottawa Bearing Dataset}
In the uOttawa dataset, experiments are performed on a SpectraQuest machinery fault simulator (MFS-PK5M). The accelerometer was placed on the housing of the experimental bearing to collect the vibration data with a sampling rate of 200,000 samples per second. The measurements have two experimental settings: bearing health condition and varying speed condition. The health conditions of the bearing include healthy, faulty with an IR defect, faulty with an OR defect, faulty with a ball defect, and faulty with combined defects on the IR, the OR, and the ball. The operating rotational speed conditions are increasing speed, decreasing speed, increasing then decreasing speed, and decreasing then increasing speed. Hence, there are twenty different settings in the measurements. Accordingly, the dataset used in this paper is arranged under one healthy class and four faulty classes so that each class includes all the five varying speed conditions as shown in Table \ref{dataset_table}. To evaluate the proposed method at different time durations of input segments, each vibration signal is divided into input segments of 5,000 and 40,000 data points. With a sampling rate of 200,000 samples per second, this gives input segments with $T_{v_{in}}$ values of 0.025 seconds and 0.2 seconds, respectively. For each time duration, values of $k=5, 7$ and $m= 5,7, 9$ are used to decompose the input segment and construct feature vectors.\\

The \textit{Daubechies 4 (dp4)} base wavelet is selected to decompose input vibration segments and reconstruct the elementary waveforms. After the feature extraction stage, the resultant datasets of features are divided into 80\% training samples and 20\% test samples. Accordingly, classifiers are trained on these datasets, and Bayesian Optimization with Gaussian Processes \cite{lyang} is used to tune the hyperparameters. Finally, test samples are used to evaluate the performance of the proposed method. In order to select a suitable classifier, the performances of three classifiers are compared on the CWRU dataset, and the classifier of best performance is selected to evaluate the performance on the three datasets. The three classifiers are Support-Vector Machines (SVM), eXtreme Gradient Boosting (XGBoost), and Random Forest (RF). Python programming language, PyWavelets, and SciPy libraries are used to build the models\footnote[1]{Code is available at: \url{https://github.com/Western-OC2-Lab/Vibration-Based-Fault-Diagnosis-with-Low-Delay}}. Results and related discussion are presented in the next section.

\begin{table*}
\caption{Experimental Datasets.}
\begin{tabular}{|c|c|c|c|c|c|c|c|}
\hline
\multicolumn{1}{|c|}{\multirow{11}[1]{*}{\textbf{CWRU dataset}}} & \textbf{Class} & \textbf{Health condition} & \textbf{Fault diameter} & \multicolumn{4}{c|}{\textbf{Motor speed (rpm)}} \\
\cline{2-8}      & 1     & Healthy & \multicolumn{1}{c|}{NA} & 1730  & 1750  & 1772  & 1797 \\
\cline{2-8}      & 2     & \multirow{3}[1]{*}{IR  faults} & 0.07" & 1730  & 1750  & 1772  & 1797 \\
\cline{2-2}\cline{4-8}      & 3     & \multicolumn{1}{c|}{} & 0.014" & 1730  & 1750  & 1772  & 1797 \\
\cline{2-2}\cline{4-8}      & 4     & \multicolumn{1}{c|}{} & 0.021" & 1730  & 1750  & 1772  & 1797 \\
\cline{2-8}      & 5     & \multirow{3}[1]{*}{Ball faults} & 0.07" & 1730  & 1750  & 1772  & 1797 \\
\cline{2-2}\cline{4-8}      & 6     & \multicolumn{1}{c|}{} & 0.014" & 1730  & 1750  & 1772  & 1797 \\
\cline{2-2}\cline{4-8}      & 7     & \multicolumn{1}{c|}{} & 0.021" & 1730  & 1750  & 1772  & 1797 \\
\cline{2-8}      & 8     & \multirow{3}[1]{*}{OR faults} & 0.07" & 1730  & 1750  & 1772  & 1797 \\
\cline{2-2}\cline{4-8}      & 9     & \multicolumn{1}{c|}{} & 0.014" & 1730  & 1750  & 1772  & 1797 \\
\cline{2-2}\cline{4-8}      & 10    & \multicolumn{1}{c|}{} & 0.021" & 1730  & 1750  & 1772  & 1797 \\
\hline
\multicolumn{1}{|c|}{\multirow{5}[1]{*}{\textbf{ PU dataset}}} & \textbf{Class} & \textbf{Health condition} & \textbf{Fault type} & \multicolumn{4}{c|}{\textbf{Motor speed (rpm)}} \\
\cline{2-8}      & 1     & Healthy & \multicolumn{1}{c|}{NA} & \multicolumn{4}{c|}{\multirow{4}[1]{*}{1500}} \\
\cline{2-4}     & 2    & Combined IR and OR faults  & Multiple damages & \multicolumn{4}{c|}{} \\
\cline{2-4}     & 3    & IR faults & Single, repetitive, and multiple damages & \multicolumn{4}{c|}{} \\
\cline{2-4}     & 4    & OR faults & Single and repetitive damages & \multicolumn{4}{c|}{} \\
\hline
\multicolumn{1}{|c|}{\multirow{6}[1]{1cm}{\textbf{uOttawa dataset}}} & \textbf{Class} & \multicolumn{2}{c|}{\textbf{Health condition}} & \multicolumn{4}{c|}{\textbf{Speed conditions}} \\
\cline{2-8}    & 1  & \multicolumn{2}{c|}{Healthy} & \multicolumn{4}{c|}{\multirow{5}{4cm}{Increasing speed, decreasing speed, increasing then decreasing speed, and decreasing then increasing speed}} \\
\cline{2-4} & 2 & \multicolumn{2}{c|}{IR faults} & \multicolumn{4}{c|}{} \\
\cline{2-4} & 3 & \multicolumn{2}{c|}{OR faults} & \multicolumn{4}{c|}{} \\
\cline{2-4} & 4 & \multicolumn{2}{c|}{Ball faults} & \multicolumn{4}{c|}{} \\
\cline{2-4} & 5 & \multicolumn{2}{c|}{Combined IR, OR, and ball faults } & \multicolumn{4}{c|}{} \\
\hline
\end{tabular}%
\centering
\label{dataset_table}
\end{table*}

\begin{figure*}[t!]
\centerline{\includegraphics[width=\textwidth]{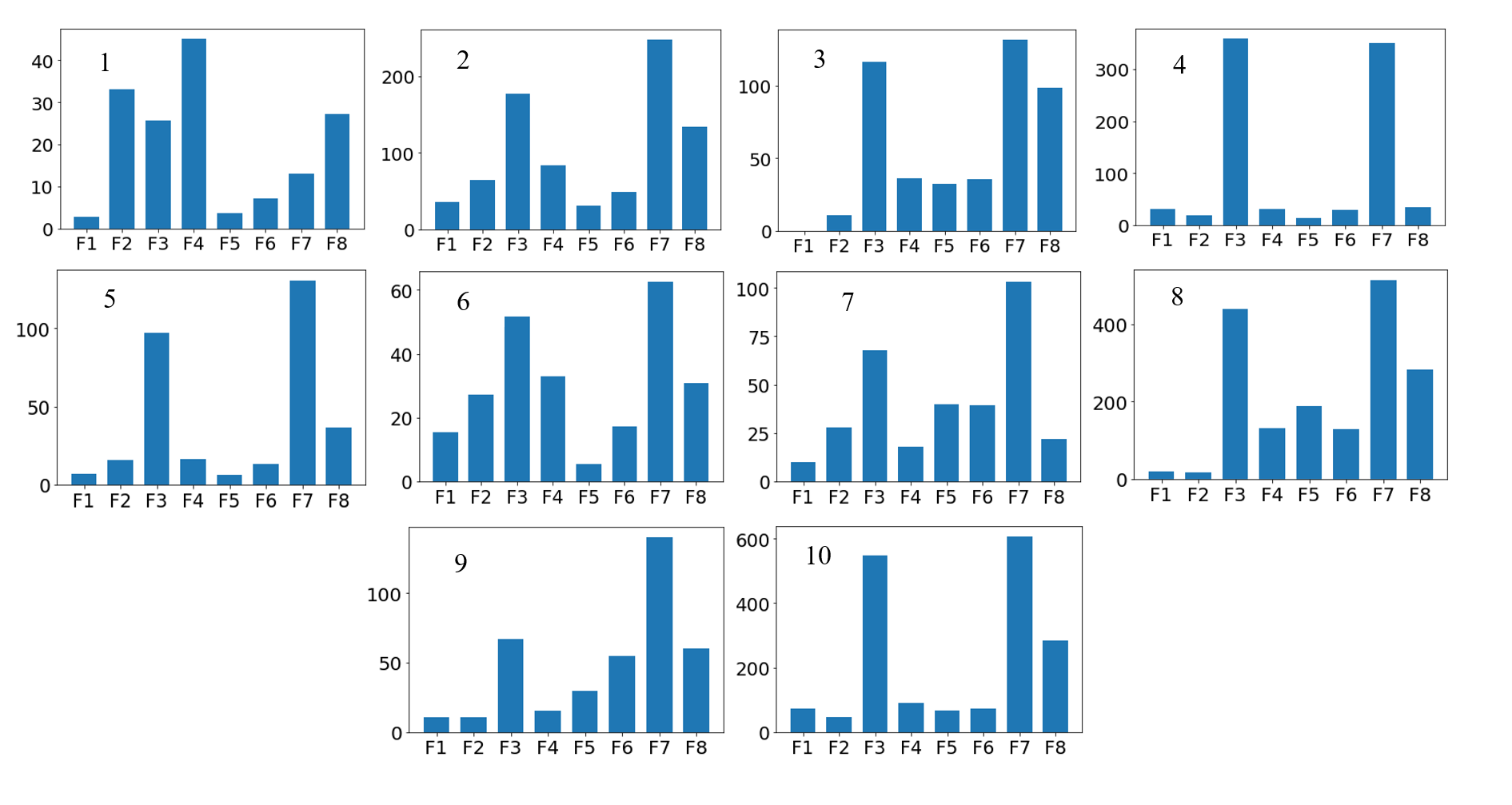}}
\caption{Feature sample from each class of CWRU dataset with $T_{v_{in}}=0.05$ seconds, $k=3$ and $m=1$.}
\label{fig3}
\centering
\end{figure*}

\renewcommand{\arraystretch}{1.7}
\begin{table*}[t!]
\caption{Performance comparison of SVM, XGBoost, and RF classifiers on CWRU dataset: Input segment duration = 0.05 seconds (600 samples), ($k=3, m=1$).}
\begin{center}
\begin{tabular}{|c|c|c|c|}
\hline
Performance Metric & Support Vector Machines (SVM) & eXtreme Gradient Boosting (XGBoost) & Random Forest (RF)\\
\hline
ROC AUC Score & 1.000 & 1.000 & 1.000\\
\hline
Accuracy (\%) & 98.013 & 98.344 & 99.172\\
\hline
\end{tabular}
\end{center}
\label{tableR1}
\end{table*}

\renewcommand{\arraystretch}{1.7}
\begin{table*}[t!]
\caption{Performance comparison between proposed features and other features on CWRU dataset: Input segment duration = 0.05 seconds (600 samples).}
\begin{center}
\begin{tabular}{|c|c|c|c|c|c|}
\hline
Performance Metric & Mean & Crest Factor & Kurtosis & Shannon Entropy & Proposed  ($k=3, m=1$)\\
\hline
ROC AUC Score & 0.999 & 0.937 & 0.963 & 1.000 & 1.000\\
\hline
Accuracy (\%) & 96.689 & 64.073 & 74.834 &  98.510 & 99.172\\
\hline
\end{tabular}
\end{center}
\label{tableR2}
\end{table*}

\begin{table*}
\caption{Performance results of the proposed method on CWRU dataset.}
\begin{tabular}{|c|c|c|c|c|c|c|c|c|c|c|c|}
\hline
\multicolumn{2}{|c|}{ Input Segment} & \multirow{3}{*}{Performance Metric} & \multicolumn{3}{c|}{\multirow{2}{0.7cm}{\textit{\textbf{k = 2}}}} & \multicolumn{3}{c|}{\multirow{2}{0.7cm}{\textit{\textbf{k = 3}}}} & \multicolumn{3}{c|}{\multirow{2}{0.7cm}{\textit{\textbf{k = 5}}}} \\
\cline{1-2}\multicolumn{1}{|c|}{\multirow{3}{0.05cm}{$N_o$}} 
& \multicolumn{1}{c|}{\multirow{3}{0.3cm}{$T_{v_{in}}$ (sec)}} &   & \multicolumn{3}{c|}{} & \multicolumn{3}{c|}{} & \multicolumn{3}{c|}{} \\
\cline{4-12}      &       &       & \textit{\textbf{m = 1}} & \textit{\textbf{m = 2}} & \textit{\textbf{m = 3}} & \textit{\textbf{m = 1}} & \textit{\textbf{m = 2}} & \textit{\textbf{m = 3}} & \textit{\textbf{m = 1}} & \textit{\textbf{m = 2}} & \textit{\textbf{m = 3}} \\
\cline{3-12}     &       & Size of feature vector $S$ & \multicolumn{1}{l|}{$1\times4$} & \multicolumn{1}{l|}{$1\times8$} & \multicolumn{1}{l|}{$1\times12$} & \multicolumn{1}{l|}{$1\times8$} & \multicolumn{1}{l|}{$1\times16$} & \multicolumn{1}{l|}{$1\times24$} & \multicolumn{1}{l|}{$1\times32$} & \multicolumn{1}{l|}{$1\times64$} & \multicolumn{1}{l|}{$1\times96$} \\
\hline
\multirow{5}{*}{300} & \multirow{5}{*}{0.025} & AUC Score & \textbf{0.991} & \textbf{0.996} & \textbf{0.996} & \textbf{0.999} & \textbf{1.000} & \textbf{1.000} & \textbf{1.000} & \textbf{1.000} & \textbf{1.000} \\
\cline{3-12}      &       & Accuracy (\%) & 89.23\% & 93.10\% & 93.05\% & 97.51\% & 98.18\% & 98.32\% & \textbf{99.63\%} & \textbf{99.11\%} & \textbf{99.14\%} \\
\cline{3-12}      &       & Online Processing Time $T_p$ (sec)* & 0.0130 & 0.0140 & 0.0140 & 0.0140 & 0.0150 & 0.0150 & 0.0170 & 0.0260 & 0.0280 \\
\cline{3-12}      &       & Peak Memory (MB)* & 0.1826 & 0.1851 & 0.1875 & 0.1889 & 0.1942 & 0.1949 & 0.2563 & 0.2598 & 0.2672 \\
\cline{3-12}      &       & System delay $\tau_d=T_{v_{in}}+T_p$ (sec) & 0.0380 & 0.0390 & 0.0390 & 0.0390 & 0.0400 & 0.0400 & 0.0420 & 0.0510 & 0.0530 \\
\hline
\multirow{5}{*}{600} & \multirow{5}{*}{0.05} & AUC Score & \textbf{0.997} & \textbf{0.999} & \textbf{0.999} & \textbf{1.000} & \textbf{1.000} & \textbf{1.000} & \textbf{1.000} & \textbf{1.000} & \textbf{1.000} \\
\cline{3-12}      &       & Accuracy (\%) & 93.54\% & 96.00\% & 96.20\% & \textbf{99.17\%} & \textbf{99.61\%} & \textbf{99.56\%} & \textbf{99.66\%} & \textbf{99.80\%} & \textbf{99.85\%} \\
\cline{3-12}      &       & Online Processing Time $T_p$ (sec)* & 0.0140 & 0.0140 & 0.0140 & 0.0150 & 0.0150 & 0.0160 & 0.0270 & 0.0289 & 0.0309 \\
\cline{3-12}      &       & Peak Memory (MB)* & 0.1897 & 0.1914 & 0.1923 & 0.1956 & 0.2019 & 0.2039 & 0.2573 & 0.2574 & 0.2687 \\
\cline{3-12}      &       & System delay $\tau_d=T_{v_{in}}+T_p$ (sec) & 0.0640 & 0.0640 & 0.0640 & 0.0650 & 0.0650 & 0.0660 & 0.0770 & 0.0789 & 0.0809 \\
\hline
\multirow{5}{*}{1,200} & \multirow{5}{*}{0.1} & AUC Score & \textbf{0.999} & \textbf{1.000} & \textbf{0.999} & \textbf{1.000} & \textbf{1.000} & \textbf{1.000} & \textbf{1.000} & \textbf{1.000} & \textbf{1.000} \\
\cline{3-12}      &       & Accuracy (\%) & 96.24\% & 99.11\% & 98.42\% & \textbf{100.00\%} & \textbf{100.00\%} & \textbf{99.80\%} & \textbf{99.80\%} & \textbf{100.00\%} & \textbf{99.31\%} \\
\cline{3-12}      &       & Online Processing Time $T_p$ (sec)* & 0.0150 & 0.0150 & 0.0150 & 0.0150 & 0.0160 & 0.0160 & \multicolumn{1}{l|}{0.02791} & 0.0299 & 0.0319 \\
\cline{3-12}      &       & Peak Memory (MB)* & 0.2148 & 0.2158 & 0.2156 & 0.2349 & 0.2353 & 0.2361 & 0.2985 & 0.3063 & 0.3098 \\
\cline{3-12}      &       & System delay $\tau_d=T_{v_{in}}+T_p$ (sec) & 0.1150 & 0.1150 & 0.1150 & 0.1150 & 0.1160 & 0.1160 & 0.1279 & 0.1299 & 0.1319 \\
\hline
\multirow{5}{*}{2,400} & \multirow{5}{*}{0.2} & AUC Score & \textbf{0.998} & \textbf{1.000} & \textbf{1.000} & \textbf{1.000} & \textbf{1.000} & \textbf{1.000} & \textbf{1.000} & \textbf{1.000} & \textbf{1.000} \\
\cline{3-12}      &       & Accuracy (\%) & 95.23\% & 99.01\% & 98.41\% & \textbf{100.00\%} & \textbf{100.00\%} & \textbf{100.00\%} & \textbf{99.80\%} & \textbf{99.83\%} & \textbf{99.80\%} \\
\cline{3-12}      &       &  Online Processing Time $T_p$ (sec)* & 0.0150 & 0.0150 & 0.0150 & 0.0155 & 0.0160 & 0.0170 & 0.0340 & 0.0349 & 0.0350 \\
\cline{3-12}      &       & Peak Memory (MB)* & 0.2796 & 0.2800 & 0.2803 & 0.3061 & 0.3066 & 0.3087 & 0.4062 & 0.4065 & 0.4089 \\
\cline{3-12}      &       & System delay $\tau_d=T_{v_{in}}+T_p$ (sec) & 0.2150 & 0.2150 & 0.2150 & 0.2155 & 0.2160 & 0.2170 & 0.2340 & 0.2349 & 0.2350 \\
\hline
\multicolumn{9}{c}{*Based on  a machine with i7-8550u CPU 1.8GHz and 8 GB RAM} \\
\end{tabular}
\label{results}
\end{table*}

\begin{table*}
\caption{Mean Accuracy (\%) on CWRU dataset.}
\begin{tabular}{|c|c|c|c|c|c|c|c|c|c|}
\hline
\multirow{2}{*}{Parameters Settings} &
\multicolumn{3}{c|}{$k=2$} &
\multicolumn{3}{c|}{$k=3$} &
\multicolumn{3}{c|}{$k=5$} \\
\cline{2-10} & $m=1$ & $m=2$ & $m=3$ & $m=1$ & $m=2$ & $m=3$ & $m=1$ & $m=2$ & $m=3$ \\
\hline
Size of feature vector $S$  & $1\times4$ & $1\times8$ & $1\times12$ & $1\times8$ & $1\times16$ & $1\times24$ & $1\times32$ & $1\times64$ & $1\times96$ \\
\hline
Mean AUC Score & 0.996 & 0.999 & 0.999 & 1.000 & 1.000 & 1.000 & 1.000 & 1.000 & 1.000 \\
\hline
Mean Accuracy (\%) & 93.56\% & 96.80\%	& 96.52\% & 99.17\% & 99.45\% & 99.42\% & 99.72\%  &  99.69\%	 &  99.53\% \\
\hline
\end{tabular}
\centering  
\label{mean_acc}
\end{table*}

\begin{figure*}
\centerline{\includegraphics[width=\textwidth]{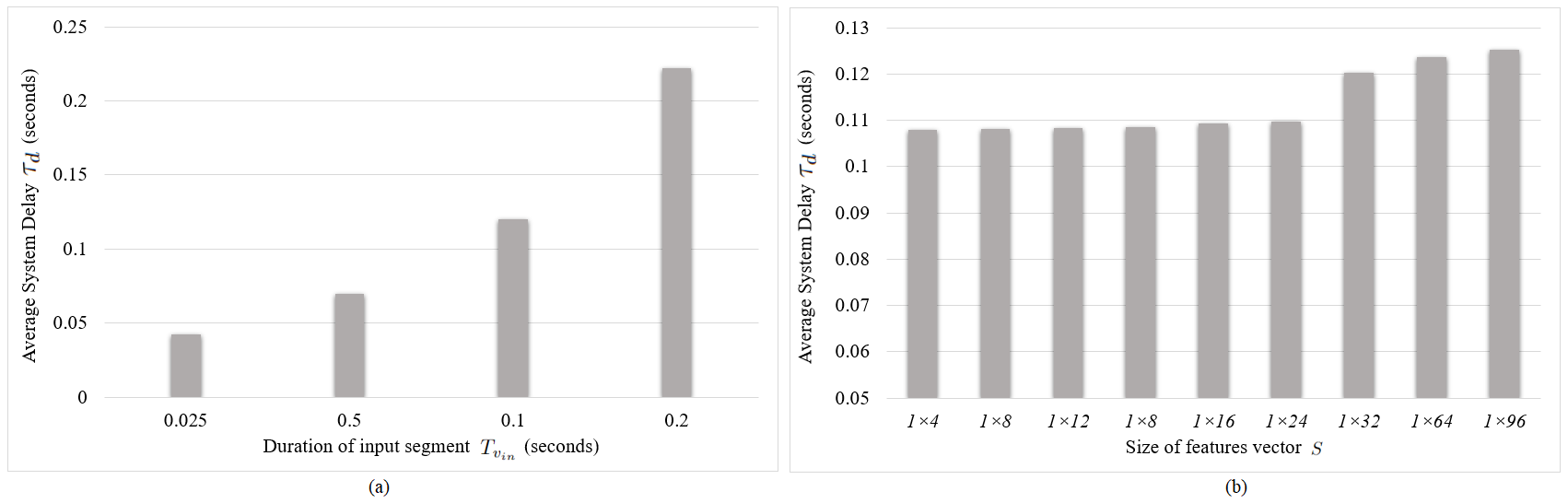}}
\caption{Average system delay $\tau_d$ as a function of (a) duration of input segment $T_{v_{in}}$ and (b) size of features vector $S$.}
\label{fig4}
\end{figure*}

\newcolumntype{L}{>{\centering\arraybackslash}m{3.5cm}}
\renewcommand{\arraystretch}{1.5}
\begin{table*}
\caption{Performance comparison between the proposed method and other methods on CWRU dataset.}
\begin{tabular}{L L L L L}
\hline
Reference & Approach & No. of data points in input vibration segment  & Size of features vector & Achieved accuracy (\%) \\
\hline
Proposed & WPD-FFT with k =5, m=1 & 300 & 1×32 & 99.63\%  \\

Proposed & WPD-FFT with  k =3, m=1 & 600 & 1×8 & 99.17\%  \\

Proposed & WPD-FFT with  k =3, m=1 & 1200 & 1×8 & 100.00\%  \\

Proposed & WPD-FFT with  k =3, m=1 & 2400 & 1×8 & 100.00\%  \\

M. Jalayer \textit{et al.}, 2021 \cite{mJalayer} & Convolutional Long Short-Term Memory (CLSTM) & 500 & \multicolumn{1}{m{3cm}}{1×((($D$($K$+3) )×$L$)+2$D$), $D=$ no.  of accelerometer sensors, $L=$ length of input segment, $K=$ first $K$ components of the raw signal acquired by CWT} & 100.00\% \\

M. Liao \textit{et al.}, 2021 \cite{mliaoc353}  & Convolutional Neural Network (CNN) with wavelet convolution and deep transfer learning & 2048 & 1×2048 & 99.73\% \\

A. Shenfield and M. Howarth, 2020 \cite{ashen} & Recurrent Neural Network (RNN) +CNN & 4096 & 1×4096 & 100.00\% \\

Y. Zhang \textit{et al.}, 2020 \cite{yzhang111} & CNN with Scaled Exponential Linear Unit (SELU) and hierarchical regularization & 512 & 64×64 & 100.00\% \\

H. Ahmed and A. K. Nandi, 2018 \cite{hahmed} & Compressive sampling and feature ranaking with ANN classifier & 2000 & 1×50 & 100.00\% \\
\end{tabular}
\centering
\label{comp}
\end{table*}

\begin{table*}
\caption{Performance results of the proposed method on the PU dataset.}
\begin{tabular}{|c|c|c|c|c|c|c|c|c|}
\hline
\multicolumn{2}{|c|}{ Input Segment} & \multirow{3}[6]{*}{Performance Metric} & \multicolumn{3}{c|}{\multirow{2}[4]{*}{\textit{\textbf{k = 3}}}} & \multicolumn{3}{c|}{\multirow{2}[4]{*}{\textit{\textbf{k = 5}}}} \\
\cline{1-2}\multicolumn{1}{|c|}{\multirow{3}[6]{*}{$N_o$}} & \multicolumn{1}{c|}{\multirow{3}[6]{*}{$T_{v_{in}}$ (sec)}} &       & \multicolumn{3}{c|}{} & \multicolumn{3}{c|}{} \\
\cline{4-9}      &       &       & m = 3 & m = 5 & m = 7 & m = 3 & m = 5 & m = 7 \\
\cline{3-9}      &       & Size of feature vector $S$ & $1\times24$ & $1\times40$ & $1\times56$ & $1\times96$ & $1\times160$ & $1\times224$ \\
\hline
\multirow{2}[4]{*}{1,600} & \multirow{2}[4]{*}{0.025} & AUC Score & 0.979 & 0.979 & 0.977 & \textbf{0.994} & \textbf{0.993} & \textbf{0.992} \\
\cline{3-9}      &       & Accuracy (\%) & 86.55\% & 86.53\% & 85.99\% & 91.91\% & 91.32\% & 90.29\% \\
\hline
\multirow{2}[4]{*}{12,800} & \multirow{2}[4]{*}{0.2} & AUC Score & \textbf{0.998} & \textbf{0.998} & \textbf{0.999} & \textbf{1.000} & \textbf{1.000} & \textbf{1.000} \\
\cline{3-9}      &       & Accuracy (\%) & 96.88\% & 96.44\% & 97.19\% & \textbf{99.56\%} & \textbf{100.00\%} & \textbf{99.69\%} \\
\hline
\end{tabular}%
\label{results_pader}
\centering
\end{table*}

\newcolumntype{J}{>{\centering\arraybackslash}m{5.5cm}}
\renewcommand{\arraystretch}{1.5}
\begin{table*}
\caption{Performance comparison between the proposed method and other methods on PU dataset.}
\begin{tabular}{J J J}
\hline
Reference & Approach & Achieved accuracy (\%) \\
\hline
Proposed & WPD-FFT with $k =5$, $m=5$, length of input segments = 12800 data points & 100.00\%  \\

D. Wang \textit{et al.}, 2021 \cite{dwangn} & Attention-based Multi-Dimensional Concatenated CNN (AMDC-CNN) & 99.80\% \\

A. K. Sharma and N. K. Verma, 2021 \cite{akq} & Deep Neural Network (DNN) with Net2Net transformation and domain adaptation  & 96.24\% \\

L. Hou \textit{et al.}, 2020 \cite{lhi} & Input Feature Mappings (IFMs)-based Deep Residual Network (ResNet) & 99.70\% \\

\end{tabular}
\centering  
 \label{comp_pu}
\end{table*}

\section{Results and Discussion}

\subsection{Classifier Selection}

As mentioned earlier, performances of SVM, XGBoost, and RF classifiers are compared on the CWRU dataset, and the classifier of best performance is selected to evaluate the performance of the proposed method on the three datasets. SVM creates a hyperplane that splits the input features space. For inputs belonging to $N$ classes, their feature space is an $N$-dimensional space representing these inputs and their associated classes. Hence, SVM attempts to create a hyperplane that achieves the best separation between input features according to their classes. XGBoost and RF are decision trees-based classifiers. They both utilize ensemble learning techniques and attempt to predict the target class by combining the estimates from individual decision trees. However, XGBoost builds a single decision tree at a time; each new tree predicts the residuals or errors of the previously trained decision tree. In contrast to XGBoost, RF fits several decision tree classifiers on various subsets of training datasets using different subsets of features. Accordingly, decisions of individual classifiers are aggregated to obtain a final decision and predict the target class. TABLE. \ref{tableR1} compares performances of the three classifiers on the CWRU dataset in terms of AUC score and accuracy. Vibration input segments have 600 data points which corresponds to 0.05 seconds. While the three classifiers achieved high and comparable performance levels, tree-based classifiers (XGBoost and RF) have slightly better performance than SVM. However, the RF classifier has a slightly better performance compared to XGBoost. The main advantage of RF over XGBoost is that RF fits several decision trees and trains each tree independently using a random subset of the data. This increases randomness and reduces bias in the training phase, allowing for better generalization. Hence, RF is more robust to overfitting compared to single decision tree-based algorithms such as XGBoost. Accordingly, RF will be used from now onwards to evaluate the performance of the proposed method on the three datasets.

\begin{table*}
\caption{Performance results of the proposed method on the uOttawa dataset.}
\begin{tabular}{|c|c|c|c|c|c|c|c|c|}
\hline
\multicolumn{2}{|c|}{ Input Segment} & \multirow{3}[6]{*}{Performance Metric} & \multicolumn{3}{c|}{\multirow{2}[4]{*}{\textit{\textbf{k = 5}}}} & \multicolumn{3}{c|}{\multirow{2}[4]{*}{\textit{\textbf{k = 7}}}} \\
\cline{1-2}\multicolumn{1}{|c|}{\multirow{3}[6]{*}{$N_o$}} & \multicolumn{1}{c|}{\multirow{3}[6]{*}{$T_{v_{in}}$ (sec)}} &       & \multicolumn{3}{c|}{} & \multicolumn{3}{c|}{} \\
\cline{4-9}      &       &       & m = 5 & m = 7 & m = 9 & m = 5 & m = 7 & m = 9 \\
\cline{3-9}      &       & Size of feature vector $S$  & $1\times160$ & $1\times224$ & $1\times288$ & $1\times640$ & $1\times896$ & $1\times1152$ \\
\hline
\multirow{2}[4]{*}{5,000} & \multirow{2}[4]{*}{0.025} & AUC Score & \textbf{0.996} & \textbf{0.995} & \textbf{0.994} & \textbf{0.998} & \textbf{0.997} & \textbf{0.997} \\
\cline{3-9}      &       & Accuracy (\%) & 94.31\% & 93.94\% & 93.54\% & 96.40\% & 95.81\% & 95.50\% \\
\hline
\multirow{2}[4]{*}{40,000} & \multirow{2}[4]{*}{0.2} & AUC Score & \textbf{0.997} & \textbf{0.997} & \textbf{0.997} & \textbf{1.000} & \textbf{0.999} & \textbf{0.999} \\
\cline{3-9}      &       & Accuracy (\%) & 96.83\% & 95.67\% & 94.67\% & \textbf{98.83\%} & \textbf{98.17\%} & 97.83\% \\
\hline
\end{tabular}%
\centering
\label{results_uoo}
\end{table*}

\newcolumntype{D}{>{\centering\arraybackslash}m{5.5cm}}
\renewcommand{\arraystretch}{1.5}
\begin{table*}
\caption{Performance comparison between the proposed method and other methods on uOttawa dataset.}
\begin{tabular}{D D D}
\hline
Reference & Approach & Achieved accuracy (\%) \\
\hline
Proposed & WPD-FFT with $k =7$, $m=5$, length of input segments = 40000 data points & 98.83\%  \\

H. Geng \textit{et al.}, 2022 \cite{hhi} & Generalized Broadband Mode Decomposition (GBMD) with Distance Evaluation Technology (DET) for feature screening & 96.67\% \\

K. Zhang \textit{et al.}, 2022 \cite{kza} & Convolutional denoising auto-encoder with Convolutional Long Short-Term Memory (CLSTM) & 97.68\% \\

D. Soother \textit{et al.}, 2020 \cite{dst} & LSTM & 77.00\% \\
\end{tabular}
\centering
 \label{comp_uoo}
\end{table*}

\subsection{Effectiveness of Proposed Features}

As mentioned earlier, the proposed method utilizes high energy concentration caused by defect-induced transient vibrations to extract the features according to Eq. \eqref{1} or Eq. \eqref{2}. For each elementary waveform $i$, the features are extracted by firstly selecting the first $m$ dominant frequencies from its corresponding vectors of FFT spectrum amplitudes $A_{FFT_i}$ and FFT spectrum frequencies $F_{A_{FFT_i}}$ where $i= 1,...,2^k$. Then, each frequency is multiplied  by its amplitude value to increase the discriminative capabilities of the extracted features. Fig.\ref{fig3} shows a sample from features vector for each class of CWRU dataset with $T_{v_{in}}= 0.05$ seconds (corresponds to 600 data points), $k=3$, and $m=1$. As shown, the features have obvious different patterns, which reflect the high discriminative degree of the extracted features.
Performance comparison on the CWRU dataset is conducted between the proposed features with $k=3$ and $m=1$  and other features extracted from vectors of FFT spectrum amplitudes $A_{FFT_i}$ to demonstrate the effectiveness of the proposed features. These features include mean, crest factor, kurtosis, and Shannon entropy. Performance comparison is shown in TABLE \ref{tableR2}. Results demonstrate the effectiveness of proposed features as they achieved the best performance among other features.

\subsection{Experimental Results on CWRU Dataset}
Table \ref{results} shows performance results of the proposed method on the CWRU dataset where accuracy, AUC score, system delay, and complexity--in terms of online processing time and memory requirements-- are used to evaluate the performance. In terms of accuracy and AUC score, the proposed method achieved very high performance  (accuracy $>99\%$, AUC score $=1.00$) with all durations of the input vibration segment. Moreover, it reached 100\% accuracy using an input duration of 0.1 seconds (1,200 data points) with 8 features only. These results reflect the high sensitivity of the proposed method to fault-related transients. In terms of the duration of the input segment, it significantly impacts the accuracy. For fixed parameters setting ($k, m$), the higher the duration, the higher the accuracy and the higher the delay. However, proper setting of the system parameters $(k, m)$ improves the system sensitivity and achieves a good trade-off between the input duration and the system accuracy. For instance, as shown in the table, with an input duration of 0.025 seconds (300 data points), setting ($k=5, m=1$) brought more than 11\% improvement in the accuracy compared to $(k=2, m=1)$. On the other hand, with 0.2 seconds (2,400 data points) of input duration, increasing the $k$ value from $3$ to $5$ has slightly affected the system accuracy. Similarly, with 0.025 seconds (300 data points) of input duration, increasing the $m$ value from $1$ to $3$ with $k=3$ slightly affected the accuracy. This slight degradation in the accuracy can be explained by increased redundancy in the extracted features. Extracting more features from vibration signals could either increase redundancy or improve sensitivity to faults in the extracted features. This generally depends on the length of the input vibration segment and fault severity. The results demonstrate that the proposed method can address this issue with high flexibility through $k$ and $m$ parameters.

Regarding the influence of system parameters on system accuracy, TABLE \ref{mean_acc} shows mean accuracy and mean AUC score as a function of system parameters ($k, m$). As results generally indicate, the decomposing level parameter $k$ has a considerable impact on system accuracy compared to the number of selected dominant frequencies $m$. More specifically, increasing the value of $k$ with fixed $m$, brought more improvement compared to increasing the value of $m$ with fixed $k$.\

Regarding system delay $\tau_d$, the proposed method achieved excellent performance (accuracy $>99\%$, AUC score $=1.00$) with a minimum system delay $\tau_d$ of 0.042 seconds. As mentioned earlier, with fixed computational resources, the system delay $\tau_d$ is a function of input segment duration $T_{v_{in}}$ and size of features vector $S$. Fig. \ref{fig4} shows empirical sensitivity analysis of average system delay as a function of (a) duration of input segment $T_{v_{in}}$ and (b) size of features vector $S$. Here, the average system delay at each value of  $T_{v_{in}}$/($S)$ is calculated by averaging individual system delays at corresponding values of $S$/($T_{v_{in}}$). The results reveal that $\tau_d$ is heavily influenced by $T_{v_{in}}$ compared to $S$. For instance, increasing the input duration from $T_{v_{in}}=0.025$ seconds to  $T_{v_{in}}=0.2$ seconds (700\% increase in input segment duration), increased the average system delay by 424.5\%. On the other hand, increasing the size of features vector from $S=1\times4$ to $S=1\times96$ (2300\% increase in features size), increased the average system delay by 15.9\% only.\

According to the analysis in Theoretical Background Section, for a given condition monitoring system and considering fixed computational resources, the system delay is influenced by the time duration of the input vibration signal (or equivalently the number of datapoint samples in the input vibration signal) and the size of the extracted features. Hence, as mentioned earlier, a reliable design of a vibration-based condition monitoring system with low system delay would minimize number of datapoint samples in the input vibration signal, minimize size of extracted features, and maximize accuracy. Thus, to demonstrate the efficiency of the proposed method in this regard, a comparison between the proposed method and other methods is conducted in terms of number of data points in the input vibration segment, size of features vector, and achieved accuracy. Table \ref{comp} shows the comparison results. The length of the input segment and the size of the feature vector are used to assess the system delay since the paper establishes the direct relationship between the system delay and the length of the input vibration segment and resultant feature size, as formulated in Theoretical Background  Section. As shown in the table, the proposed method reached 100\% accuracy using an input segment of 1,200 data points with 8 features only. On the other hand, other methods require either longer segments of the input signal \cite{ashen} or use more features \cite{hahmed}\cite{mJalayer}\cite{ashen}\cite{yzhang111} compared to our proposed method to reach 100\% accuracy. Moreover, the proposed method does not involve extra computations for feature ranking or dimensionality reduction compared to other methods.

\subsection{Experimental Results on PU Dataset}
The performance of the proposed method is further validated on the PU dataset, which includes real accelerated-lifetime damages. Performance results are shown Table \ref{results_pader}. As shown, the proposed method achieved an excellent accuracy of 99.56\% by extracting 96 ($k=5, k=3$) features from input vibration segments of 0.1 seconds duration (12,800 data points). Furthermore, it reached 100\% accuracy and a 1.00 AUC score by extracting 160 features ($k=5, m=5$) only. On the other hand, extracting more features (224 features using $k=5$ and $m=7$) led to a slight degradation in the accuracy. These results demonstrate the flexibility of the proposed method in tuning the size of extracted features which, in turn, lessens feature redundancy, improves the accuracy, and reduces the complexity. Table \ref{comp_pu} provides a comparison between the proposed method and some recent works \cite{dwangn}-\cite{lhi} on the PU dataset of healthy bearings and accelerated-lifetime damages. Results show the superiority of the proposed method in terms of achieved accuracy. Regarding the used approach, the proposed method relies completely on signal processing (WPT and FFT) to extract the features, while other methods utilize Deep Neural Networks (DNN) for feature extraction. Hence, the proposed method involves less training and tuning complexity than the other methods.

\subsection{Experimental Results on uOttawa Dataset}
To further validate the effectiveness of the proposed method under varying motor speeds, its performance is evaluated on the uOttawa dataset. Table \ref{results_uoo} shows performance results where the proposed method achieved excellent performance of 98.83\%  accuracy and 0.999 AUC score with input vibration durations of 0.1 seconds (40,000 data points) and 640 features ($k=7, m=5$). Table \ref{comp_uoo} shows a comparison between the proposed method and some of recent works \cite{hhi}-\cite{dst} on the uOttawa dataset. The proposed method achieved the best accuracy among other methods, as shown in the table. These results, along with the previous results obtained on CWRU and PU datasets, demonstrate the effectiveness of the proposed method and affirm its capability of achieving performance requirements under actual operational conditions. Furthermore, the results confirm the flexibility of the proposed method in adapting the size of extracted features according to specific application requirements.

\section{Conclusion}
In this paper, we defined and analyzed the end-to-end delay of vibration-based condition monitoring systems and introduced the concept of system delay to assess it. With fixed computational resources, the system delay depends entirely on the duration of the input vibration segment, computation steps of the used algorithm/s, and the size of the features vector. Accordingly, a low-system delay method is introduced for vibration-based condition monitoring and fault diagnosis of rolling bearings. The proposed method uses a hybrid WPT-FFT approach where it decomposes small durations of the input vibration signal using $k$-level WPT  into $2^k$ elementary waveforms with high time-frequency localization. Then it obtains the spectral components of these waveforms using FFT. Accordingly, the proposed method utilizes the first $m$ dominant frequency components in the spectrum of each waveform to construct a small number of features with high sensitivity to fault-related transients. Hence, the proposed method has a high sensitivity to fault-related transients with relatively short durations of input vibration segments. Moreover, it allows controlling the size of extracted features through ($k, m$) settings, which helps to reduce redundancy and provides high flexibility to customize the proposed method according to specific application requirements. The proposed method has been evaluated on three datasets using different durations of input vibration segments and with different parameter settings. The experimental results show that the proposed method can achieve excellent accuracy in fault diagnosis with low system delay even with combined defects and under varying motor speeds.

\section*{Acknowledgment}

This work was funded in part by National Research Council Canada under Project no.: AM-105-1.

\ifCLASSOPTIONcaptionsoff
  \newpage
\fi

\end{document}